# Haircutting Non-cash Collateral

Wujiang Lou[1]

HSBC

1st Draft February, 2016; Revised April 8, 2017.

**Abstract**

Haircutting non-cash collateral has become a key element of the post-crisis reform of the shadow banking system and OTC derivatives markets. This article develops a parametric haircut model by expanding haircut definitions beyond the traditional value-at-risk measure and employing a double-exponential jump-diffusion model for collateral market risk. Haircuts are solved to target credit risk measurements, including probability of default, expected loss or unexpected loss criteria. Comparing to data-driven approach typically run on proxy data series, the model enables sensitivity analysis and stress test, captures market liquidity risk, allows idiosyncratic risk adjustments, and incorporates relevant market information. Computational results for main equities, securitization, and corporate bonds show potential for uses in collateral agreements, e.g. CSAs, and for regulatory capital calculations.

---

[1] *The views and opinions expressed herein are the views and opinions of the author, and do not reflect those of his employer and any of its affiliates.*

## 1. Introduction

Haircut, a discount on the market value of securities taken in as collateral, draws its intuition from earlier stock loan brokers' desire to withstand stock market meltdowns without losses. That intuition remains largely intact, although when statisticians got involved and historical data are abundant, a confidence interval was used to qualify the haircut. For example, a 15% haircut would give 99% confidence of no loss within 10 days. Or equivalently in the value-at-risk (VaR) language, one is 99% confident that the stock price won't decline by more than 15% in 10 days. Naturally, a simple intuition like this does not call for sophisticated methodologies or models. Today, haircuts are determined either by rule of thumbs, subject to negotiations, or via regulatory fixings, depending on purposes. Standard financial transaction documents typically typeset negotiated and agreed haircut levels, as a governing element in the MRA (Master Repurchase Agreement) for repos, CSA (Credit Support Annex to ISDA Master Agreement) for swaps and derivatives, and exchange or central counterparty (CCP) clearing agreements. Bank for International Settlements (BIS)'s BASEL risk capital framework is the only place where the term "haircut model" is found, in addition to its schedule of standardized supervisory market price volatility haircuts. For advanced banks adopting BASEL's market risk capital rules, VaR methodology with at least 2 year historical data and internal haircut models[2] are allowed, although no technical specifics are given.

BASEL's haircut models and FSB's enhanced haircut framework (FSB 2015) exemplify a data-driven approach to haircuts, with stated qualitative and quantitative standards to guide prudential practices. Obviously, a data centric approach is as good as data is, and carries the usual caveat that history may or may not repeat itself. Except for some on-the-run government securities, debt instruments do not possess market liquidity anywhere close to the equity market. It is customary in practice that historical data are sourced from a proxy index or a representative portfolio that are close or similar in key product design features and risk characteristics, such as credit rating and maturity. For example, single 'A' rated, 3 to 5 year remaining maturity US corporate bonds issued by the financial firms could form a group. Broker/dealers then select relatively liquid ones to form an index and compute a daily index level. Some indices are constructed more rigorously and transparently than others. As underlying bonds trade sparsely, not necessarily daily, the determination of daily index levels takes both experience and good feel. So data accuracy is not assured. Another pitfall of the proxy data is that it eradicates idiosyncrasy in its totality. Imagine when a bond is on a downgrade watch or has seen large spread widening relative to its peers, one would suspect it's riskier than others and its haircut would deviate away and move higher.

---

[2] FSB's new framework strengthens BASEL 3 by establishing a haircut floor for non-centrally cleared securities financing transactions and requiring use of at least 5 year's historical data including at least one stress period (FSB 2015). Perhaps because of their internal nature, there are no published haircut models. Our general understanding is that banks' internal models are basically methods of selecting and justifying a proxy index or portfolio appropriate for collateral asset classes or subclasses.

In yet another situation, historical price or spread data could be available but not for sufficient length to experience a stress period. The unprecedented price behavior of US residential mortgage backed securities (RMBS) in 2007~2008, for instance, rendered its prior pricing history meaningless for setting haircuts via VaR. In particular, investment grade (IG) RMBS on subprime mortgage loans of 2005~2007 vintages had priced close to par before plunging to teens deep in the crisis. The VaR estimate at the time would have predicted single digit haircuts for AA- or higher rated securitization debts, in line with BASEL II's 8% haircut for corporate and securitization debts rated AAA to AA- with residual maturity greater than 5 years. Some banks, however, promptly hiked up bilateral repo haircuts in multiples (Gorton and Metrick 2012), fearful of future price volatilities which were exhibited already in their synthetic kin – ABX[3]. By relying exclusive and directly on historical data, data driven haircut methodologies lack a way of incorporating useful information and projections when they become available.

In situations like these, a parametric haircut model that builds in asset volatility and measures relating to credit quality could offer some relief. A mathematical model of haircuts, however, has been absent in the literature, probably because of VaR methodology' dominance and perceived simplicity. This article constitutes a first effort to develop a parametric haircut model from the asset pricing and credit risk perspectives. It contributes to the literature by introducing credit risk measures such as probability of default (PD), expected loss (EL), and unexpected loss (UL), to define haircuts such that these measures could satisfy certain predetermined criteria, e.g., 'AAA' rating. Essentially, the original intuition of loss aversion is transformed to the credit enhancement language typical of credit derivatives and structured products. Haircut is thus treated as a credit enhancement tool that could play its role in conjunction with other tools, such as borrower credit supports and tranching. Repo style transactions, for instance, have recourse to the borrower's general credit and as such repo haircuts are typically counterparty dependent. Haircuts in CSAs and regulatory capital context are counterparty independent as collateral is purposely used to mitigate counterparty credit risk. The extended definitions allow counterparty independent haircut and counterparty dependent haircut to be modeled within the same analytical framework. This paper aims at developing a counterparty independent haircut model capturing asset volatility, jumps and market liquidity risk. Counterparty dependent haircuts are dealt with separately (Lou 2016b).

---

[3] In 2007-2008, ABX.HE, an index of CDS on subprime RMBS bonds, is much more liquid than its referenced cash bonds. A large pricing disparity existed (Lou 2009) that aroused a vigorous debate about whether it's due to excessive speculation in the synthetic market or the cash market's failure in projecting future subprime losses. The subsequent housing market crash made a new history and showed the fortunes of the cash RMBS.

The parametric haircut model necessitates estimation or calibration of the underlying asset's jump-diffusion model parameters. Ideally, such a model should be estimated from a reliable historical data or proxy. Then idiosyncratic factors and useful market information can be incorporated through shifts in one or few relevant model parameters. It could complement existing data-driven models and could find applications in haircut design for haircutting non-cash collateral assets admitted in CSAs and exchanges or CCPs' clearing and margining agreements. It can also be used as a candidate regulatory internal haircut model, following the Financial Stability Board (FSB)'s final document on the strengthened regulatory haircut framework, issued in 2015 and expected to be adopted into BASEL and implemented by the end of 2018.

**2. Haircut Definitions**

In a repo-style securities financing transaction, the repo buyer or lender is exposed to the borrower's default risk for the whole duration with a market contingent exposure, framed on a short window for default settlement. A margin period of risk (MPR) is a time period starting from the last date when margin is met to the date when the defaulting counterparty is closed out with completion of collateral asset disposal. MPR could cover a number of events or processes (Andersen, Pykhtin and Sokol 2016), including collateral valuation, margin calculation, margin call, valuation dispute and resolution, default notification and default grace period, and finally time to sell collateral to recover the lent principal and accrued interest. If the sales proceeds are not sufficient, the deficiency could be made a claim to the borrower's estate, unless the repo is non-recourse. The lender's exposure in a repo during the MPR is simply principal plus accrued and unpaid interest. Since the accrued and unpaid interest is usually margined at cash, repo exposure in the MPR is flat.

A flat exposure could apply to OTC derivatives as well. For an OTC netting set under CSA, the mark-to-market of the derivatives could fluctuate as its underlying prices move. The derivatives exposure is formally set on the early termination date which could be days behind the point of default. The surviving counterparty, however, could have delta hedged against market factors following the default so that the derivative exposure remains a more manageable gamma exposure. As we aim at developing a collateral haircut model, we assume a constant exposure during the MPR.

It is well understood that the primary driver of haircuts is asset volatility. Market liquidity risk is another significant one, as liquidation of the collateral assets might negatively impact the market, if the collateral portfolio is illiquid, large, or concentrated in certain asset sectors or classes. Market prices could be depressed, bid/ask spreads could widen, and some assets might have to be sold at a steep discount. This is particularly pronounced with private securitization and lower grade corporates, which trade infrequently and often rely on valuation services rather than actual market quotations[4]. A haircut model therefore needs to capture liquidity risk, in addition to asset volatility.

In an idealized setting, we therefore consider a counterparty (or borrower) C's default time at $t$, when the margin is last met, an MPR of $u$ during which there is no margin posting, and the collateral assets are sold at time $t+u$ instantaneously at the market, with a possible liquidation discount $g$.

Denote the collateral market value as $B(t)$, exposure to the defaulting counterparty C as $E(t)$. At time $t$, one share of the asset is margined properly, i.e., $E(t)=(1-h)B(t)$, where $h$ is a constant haircut, $1>h \geq 0$. The margin agreement is assumed to have a zero minimum transfer amount. The lender would have a residual exposure $(E(t)-B(t+u)(1-g))^+$, where $g$ is a constant, $1>g\geq 0$. Exposure to C is assumed flat after $t$. We can write the loss function from holding the collateral as follows,

---

[4] A sample quote for CMBX.9 'AA' bond on 3/4/2016 is 93.25-bid and 94.75-ask, a spread of 1.5 point, roughly 2%. During the height of the financial crisis, bid/ask spreads of investment grade CMBS and RMBS are frequently seen in 3+ points while prices were in the 60's, roughly a 5% stress liquidity discount.

$$L(t + u) = E_t \left(1 - \frac{B_{t+u}}{B_t} \frac{1-g}{1-h}\right)^+ = (1 - g) B_t \left(1 - \frac{B_{t+u}}{B_t} - \frac{h-g}{1-g}\right)^+ \quad (1)$$

Conditional on default happening at time $t$, the above determines a one-period loss distribution driven by asset price return $B(t+u)/B(t)$. Note that for repos, this loss function is slightly different from the lender's ultimate loss which would be lessened due to a claim and recovery process. In the CSA and regulatory context, haircut is viewed as a mitigation to counterparty exposure and made independent of counterparty, so recovery from the defaulting party is not considered.

Let $y = 1 - \frac{B_{t+u}}{B_t}$ be the price decline. If $g=0$, $Pr(y>h)$ equals to $Pr(L(u)>0)$. There is no loss, if the price decline is less or equal to $h$. A first dollar loss will occur only if $y>h$. $h$ thus provides a cushion before a loss is incurred. Given a target rating class's default probability $p$, the first loss haircut can be written as

$$h_p = \inf\{h > 0 : Pr(L(u) > 0) \leq p\}. \quad (2)$$

Let $VaR_q$ denote the VaR of holding the asset, an amount which the price decline won't exceed, given a confidence interval of $q$, say $99\%$. In light of the adoption of the expected shortfall (ES) in BASEL IV's new market risk capital standard, we can define haircut as ES under the $q$-quantile,

$$h_{ES} = ES_q = E[y|y > VaR_q].$$
$$VaR_q = \inf\{y_0 > 0 : Pr(y > y_0) \leq 1 - q\}, \quad (3)$$

Without the liquidity discount, $h_p$ is the same as $VaR_q$. If haircuts are set to $VaR_q$ or $h_{ES}$, the market risk capital for holding the asset for the given MPR, defined as a multiple of VaR or ES, is zero. This implies that we can define a haircut to meet a minimum economic capital (EC) requirement $C_0$,

$$h_{EC} = \inf\{h \in R^+ : EC[L|h] \leq C_0\}, \quad (4)$$

where EC is measured either as VaR or ES subtracted by expected loss (EL). For rating criteria employing EL based target per rating class, we could introduce one more definition of haircuts based on EL target $L_0$,

$$h_{EL} = \inf\{h \in R^+ : E[L|h] \leq L_0\}, \tag{5}$$

The expected loss target $L_0$ can be set based on EL criteria of certain designated high credit rating, whether bank internal or external. With an external rating such as Moody's, for example, a firm can set the haircut to a level such that the expected (cumulative) loss satisfies the expected loss tolerance $L_0$ of some predetermined Moody's rating target, e.g., `Aaa` or 'Aa1'. In equations (4) and (5), loss $L$'s holding period does not have to be an MPR. In fact, these two definitions apply to the general trading book credit risk capital approach where the standard horizon is one year with a 99.9% confidence interval for default risk.

Different from $VaR_q$, definitions $hp$, $h_{EL}$, and $h_{EC}$ are based on a loss distribution solely generated by collateral market risk exposure. As such, we no longer apply the usual wholesale credit risk terminology of probability of default (PD) and loss given default (LGD) to determine EL as product of PD and LGD. Here EL is directly computed from a loss distribution originated from market risk and the haircut intends to be wholesale counterparty independent. For real repo transactions where repo haircuts are known to be counterparty dependent, these definitions remain fit, when the loss distribution incorporates the counterparty credit quality, as shown in Lou (2016b).

3. Collateral Price Dynamics

The loss function (equation 1) is an out-of-the-money put on the collateral asset which is predominantly decided by asset return's skewness and tail characteristics. Asset price models with stochastic volatility and jumps in both return and volatility are shown to improve empirical studies of stock indices (Eraker, Johannes, and Polson, 2003). The double exponential jump-diffusion model (DEJD, Kou 2002) is popular in exotic and path dependent options pricing, due to its appealing asymmetric jump specification and ease of transform analytics. Its extension, the mixed-exponential jump-diffusion model (MEM, Cai and Kou 2011), is capable of producing a wide variety of skewed tail distributions. Because the risk exposure window (the MPR) is a short period of time, a matter of days or weeks, stochastic volatility models are expected to have limited impact on haircut design, and could be left for future study. We choose for this exercise the DEJD and MEM, with the view that MEM could be valuable in coping with the excessive skewness and fat tails in securitization debts. Needless to say, MEM has numerical Laplace inversion procedures with error controls (Cai, Kou, and Liu 2014) and DEJD has explicit density functions (Ramezani and Zeng 2007) to allow maximum likelihood estimation.

The log return of a jump-diffusion asset price $B_t$ has the form of

$$X_t = \log\left(\frac{B_t}{B_0}\right) = \mu t + \sigma_a W_t + \sum_{j=1}^{N_t} Y_j, \tag{6}$$

where $\mu$ the asset return, $\sigma_a$ is the asset volatility, $W(t)$ a Brownian motion, $N(t)$ a Poisson process with intensity $\lambda$, and $Y_j$ a random variable denoting the magnitude of the *j*-th jump. With MEM, $Y_j$, *j=1, 2, ...*, are a sequence of independent and identically distributed mixed-exponential random variables with the pdf $f_Y(x)$ given by

$$f_Y(x) = p_u \sum_{l=1}^{m} p_l \eta_l e^{-\eta_l x} I\{x \geq 0\} + q_d \sum_{j=1}^{n} q_j \theta_j e^{\theta_j x} I\{x < 0\} \tag{7}$$

where $p_u$ and $q_d$ are up jump and down jump switching probabilities, $p_u+q_d=1$. $p_l$ is the weight (not necessarily in probability sense) of the *l*-th up jump mixture exponentially distributed at a rate of $\eta_l>1$, $\Sigma p_l=1$. Similarly $\theta_j>0$ is *j*-th down jump mixture's rate, $q_j$ weights that sum to 1, $\Sigma q_j=1$. Obviously it reduces to DEJD when *m=n=1*.

The probability of the cumulative loss $L(u) \geq b$, i.e., the tail cumulative density function (cdf), can be mapped to $X_u$'s cdf,

$$P_b|h = E[I\{L(u) \geq b\}] = Pr(X_u \leq \log(\frac{1-h-b/B_0}{1-g})) \tag{8}$$

Fixing a haircut *h*, this gives loss distribution $P_b$ as a function of *b*. Fixing *b*, $P_b$ becomes a function of *h* which can be inverted to solve for *h* given a target level of $P_b$. *VaR* can be solved by setting $P_b=1-q$. Obviously setting *b* to zero leads to equation (3). It is useful for implementations to note that equation (8) is translational in *h* and *b*, i.e., $P_b|h= P_{b*}|h^*$ where $b^*=b+(h-h^*)B_0$.

The expected loss relates to the undiscounted European put option fair value,

$$E[L(u)] = (1-g)E[(K-B_u)^+] = (1-g)P(K) \qquad (9)$$

where $K = \frac{1-h}{1-g} B_0$ and $P(K)$ is the undiscounted put fair value. Fixing $h$ thus strike $K$, the put fair value can be obtained by means of inverse Laplace transform. Finding $h$ given an EL target is a simple numerical inversion.

The intuition that a haircut is a cushion to investment loss is precisely captured in equation (9), seen easily when $L = (K - B_u)^+$ term is reformatted as $((B_0 - B_u) - hB_0)^+$ with $g=0$. If we consider an investment of $B_0$ amount, haircut $h$ effectively cuts the investment into two tranches, a senior tranche of $(1-h)B_0$ and a subordinate of $hB_0$. In the standard CDO (collateralized debt obligation) terminology, $h$ is the attachment point of the senior piece whose loss function is $L$. Different levels of haircuts then create senior/subordinate structures with varying credit enhancements. Suppose that $L_a$ and $L_b$ correspond to $h_a$ and $h_b$, $h_a<h_b$, then $L_a \geq L_b$. $L_{a,b} = L_a - L_b$ is the loss of a mezzanine tranche with attachment point of $h_a$ and detachment point of $h_b$. Such a viewpoint is relevant for securitization products where different tranches are traded in the same market, but a data-driven haircut approach would determine haircuts irrespective of their structural linkage. In particular, for market value CLO (collateralized loan obligation) where the underlying loans are actively traded, consistency between loan volatility and CLO tranche volatility becomes an important concern that will impact proper specification of tranche haircuts, a topic to be explored in the future.

Jump-diffusion models such as DEJD have been studied for stock returns primarily and bond yields. In our haircut model, it is also used for bond price returns. Modeling the price rather than yield term structure for fixed income securities can be justified because the MPR is a very short period of time. Bond options, for example, are priced using the Black-Scholes option pricing model with a log-normal bond price dynamics as they are typically of three month or shorter terms.

The conditioned loss distribution (eqt 8) and expected loss (eqt 9) do not enlist the derivatives counterparty or repo borrower's credit quality, as stated earlier these intend to capture counterparty independent haircuts. For security financing trades where counterparties or borrowers' credit strength is part of pricing consideration and repo haircuts are pro-cyclic and counterparty dependent (Gorton and Metrick 2012), the diffusion component of the DEJD model is made correlated to the dynamic spread of the counterparty or repo borrower to capture the wrong way risk, as shown in a companion paper (Lou 2016b).

## 4. Numerical techniques

Computation of loss distribution and the expected loss reduces to the cdf of $X_t$ (the log return of the jump-diffusion asset price) and the undiscounted European put valuation. These are conducted through the Laplace transform.

### 4.1 Laplacians

For a mixed exponential jump-diffusion process $X(t)$ specified in equations (6&7), Cai et al (2014) develops a two-sided Laplace transform analysis. The Laplacian for the probability density function $f_{X(t)}$ is denoted by $L_{f_{X(t)}}$,

$$L_{f_{X(t)}}(s) = E[e^{-sX_t}] = \int_{-\infty}^{\infty} e^{-sx} f_{X(t)}(x) dx = e^{G(-s)t} \tag{10}$$

where the Levy exponent function is,

$$G(x) = \frac{1}{2}\sigma_a^2 x^2 + \mu x + \lambda(p_u \sum_{l=1}^{m} \frac{p_l \eta_l}{\eta_l - x} + q_d \sum_{j=1}^{n} \frac{q_j \theta_j}{\theta_j + x} - 1) \tag{11}$$

with the range of absolute convergence (ROAC) of $(-min(\eta_l), min(\theta_j))$ for $Re(s)$, the real part of complex number $s$. The Laplace transform for the cdf $F_{X(t)}$ of $X(t)$ is then simply $L_F(s) = \frac{L_f(s)}{s}$, with its ROAC $Re(s) \in (0, min(\theta_j))$. For the undiscounted European put option with strike K, denote its fair value $P$ as a function of the normalized log strike $k$, such that $K=B_0 e^{-k}$, then $P(k)=E[(B_0 e^{-k} - B_t)^+]$. We apply the two-sided Laplace transform to $P(k)$, $k \in (-\infty, \infty)$,

$$L_{P(k)}(s) = \int_{-\infty}^{\infty} e^{-sk} E[B_0(e^{-k} - e^{X(t)})^+] dk = \frac{B_0}{s(s+1)} L_f(-s-1) \tag{12}$$

with a ROAC of $(-min(\theta_j)-1, -1)$. The Laplacian for a European call option is exactly the same as that of the put option, although the ROAC will be $Re(s) \in (0, min(\eta_j)-1)$. Note the formulae above is different from Cai et al (2014) where the Laplacian of the European call option price is conducted on the log of the strike $K$ and $L_{EuC}(s) = \frac{B_0^{s+1}}{s(s+1)} L_f(-s-1)$. Our strike normalization with moneyness is necessary as it allows use of the same error control over the put payoff (loss) and the tail *cdf* under the same parameters.

Having computed the two-sided Laplace transform of the pdf $L_{f_{X(t)}}$, cdf $L_{F_{X(t)}}$, and the loss or put $L_{P_{(k)}}$, the two sided Laplacian inversion algorithm shown in the Appendix is used to solve for loss distribution (equation 8) and expected loss (equation 9).

**4.2 Model parameters**

The DEJD model has six parameters -- ($\mu$, $\sigma$, $\lambda$, $p$, $\eta$, $\theta$) that need to be determined either through model estimation from historical data or calibration to traded instruments. In the context of regulatory haircuts, a historical estimation is generally required. As discussed earlier, historical data may not be reliably available, or are available only to a limited period of time without having experienced any stress period. In circumstances like these, calibrating the model, for example, to the options market becomes a viable choice. Especially, short maturity, deep-out-of-the money puts capture the tail behavior of the asset price dynamics. Calibration is quite standard, typically formulated as a non-linear least-square optimization problem on a chosen set of options. A risk-free discount factor can be inserted into equation (12) to arrive at the put option price. European call option prices can be obtained similarly or via put-call parity. In fact, Cai-Kou (2011) calibrates an MEM model for SPX 500 index to SPX European options. We will not repeat the calibration exercise, although the calibrated model's impact on haircuts will be discussed in the numerical example section.

When historical asset price data, either directly or as a proxy, is satisfactorily available and reliable, the model can be estimated via the maximum likelihood estimation (MLE). Ramezani and Zeng (2007) provides explicit likelihood function formula for the DEJD model and finds that DEJD produces better fit for stock indices, e.g., S&P 500 and NASDAQ composite, than for some individual stocks. This is encouraging, since, for haircut purposes, we mostly use indices or proxy portfolios returns. It, however, seems to support our concern that idiosyncratic risk factors might be overlooked in a proxy data-driven approach.

The likelihood function is written as $H(\mu, \sigma_a, \lambda, p, \eta, \theta | x) = f_{X(t)}(x | \mu, \sigma_a, \lambda, p, \eta, \theta)$ for a return data point $x$. The two-sided Laplacian of the *pdf* (equation 10) can be inverted to arrive at $H(./x)$. Ramezani and Zeng (2007) takes a different but equivalent configuration in writing that $H(\mu, \sigma_a, \lambda_u, \lambda_d, \eta, \theta | x)$ where $\lambda_u = \lambda p$ and $\lambda_d = \lambda(1-p)$ and derives a formulae involving sum of double infinite series as a result of conditioning on up and down jump counts. Since both the Laplacian inversion and the infinite series sum involve numerical truncations of infinite series and/or integrals, it's more of

an implementation efficiency choice, once both are subject to the same error bounds. Given a log return time series *{x_i}, i=1,2, …, N*, our estimation becomes an optimization problem,

$$max \prod_i H(\mu, \sigma_a, \lambda, p, \eta, \theta \mid x_i) \tag{13}$$

$s.t.\ \sigma_a > 0, \lambda > 0, 0 \leq p \leq 1, \eta > 1, \theta > 0.$

There is no other constraints than these simple lower or upper bounds. For the results presented in the next section, we have taken optimization routines off shelves from Matlab's median size constrained non-linear optimizer *fmincon*, utilizing an active-set algorithm, with sequential quadratic programming (SQP) and quasi-Newton line search.

For this study, we follow a simple estimation procedure: first calculate the sample's mean $\mu_0$ and standard deviation $\sigma_0$, then estimate the remaining four parameters of the DEJD model with ($\mu$, $\sigma_a$) fixed at ($\mu_0$, $\sigma_0$), now re-estimate the model with $\mu$ fixed at $\mu_0$, $\sigma_a$ relaxed so that five parameters are sought with initial values taken from previous estimation's outcomes, lastly repeat estimation with $\mu$ also relaxed to estimate the full set of six parameters. This procedure's intention is to find a stable local minimum close to the lognormal model, recognizing that finding the global minimum is difficult and time consuming for this type of highly non-linear optimization problem.

Comparing with a data-driven haircut approach, a parametric model of haircuts introduces potential model risk. In our context, this is related to the model parameter estimation errors or stability. We will compare different configurations of the model, perform model stability test, and conduct estimation sensitivity analysis and amend the haircut outputs with additional haircut surcharge, should the haircut's sensitivity to the estimation procedure prove to be significant.

## 5. Applications and Results

The haircut model outlined above can be used to determine repo haircuts with non-recourse or with recourse to a hindered or extremely weak repo counterparty. It can be a candidate for internal haircut models per BASEL and FSB's requirements or used to design non-cash collateral haircuts in collateral and margining agreements. Margining with collateral has become a central piece in the post-crisis redesign of the OTC derivatives markets, as seen in ISDA CSA, CCP clearing agreements, and the BCBS-IOSCO non-centrally cleared OTC derivatives margin requirements by Basel Committee on Banking Supervision (BCBS) and International Organization of Securities Commissions (IOSCO). Eligible collateral consists of, in general, cash or cash equivalents (e.g., treasury bills), government or agency debts, high grade corporate or municipal debts, equities, and asset backed securities or securitizations. Non-cash collateral consists of 25% of total collateral (ISDA 2015), among which 15% is non-government debts. ISDA also estimates that over 90% of non-centrally cleared OTC bilateral transactions would be subject to collateral agreement per BCBS-IOSCO's margin requirements, which will see higher proportion of non-government securities used as collateral.

As the main design goal is to mitigate counterparty credit risk, collateral haircuts for these non-cash collateral are necessarily counterparty independent. Such a design makes perfect sense if one thinks about the application of OIS discounting to fully *cash* collateralized derivatives trades. Such trades are indifferent to counterparties or free of bearers. Now suppose that a party wishes to substitute with non-cash collateral, say US corporate bonds, then the haircut applied should be such that it reproduces the counterparty free bearer form as close as possible. The haircut must not be counterparty dependent, or the OIS discounting shall no longer apply.

Below we demonstrate how the model is applied to three main non-cash collateral classes, equity, corporate bonds, and securitization, intertwined with considerations of liquidity risk, haircut sensitivities, and model risk.

**5.1. Equity main index**

The first data series examined is S&P 500 index (SPX), commonly used as a proxy for US main equities. To satisfy FSB's requirement of a 5 year history consisting of at least one stress period, we choose the period from 1/2/2008 to 1/2/2013 when SPX had a significant stress in the second half of 2008 and early 2009, at the height of the financial crisis. Table 1 lists three estimation results. In four parameter estimation '4-p', volatility is fixed at 26.25%, taken from the sample standard deviation, in addition to fixed µ at 0.21%. In '5-p', µ remains at 0.21%, while the rest 5 model parameters are estimated. In '6-p', all 6 DEJD model parameters are estimated. µ is large in '6-p' but the corresponding model mean is 5.87% per annum, reasonably reflecting the 5 year period's substantial equity recovery after the bust in the US housing market in 2007 and the broad financial market stress in 2008. The last two columns show the models' skewness and kurtosis. Judging from captures of sample skewness of -0.2443 and kurtosis or 9.95, estimation '4-p' is bad, '5-p' and '6-p' are pretty good.

Table 1. Estimated DEJD Model Parameters for S&P 500 index daily return from 1/2/2008 to 1/2/2013, with sample skewness of -0.2443 and kurtosis of 9.95.

|     | μ | σ | $\lambda_u$ | $\lambda_d$ | $\eta_u$ | $\eta_d$ | skewness | kurtosis |
|-----|---|---|---|---|---|---|---|---|
| 4-p | 0.0021 | 0.2625 | 19.97 | 29.85 | 103.83 | 105.74 | -0.0323 | 3.41 |
| 5-p | 0.0021 | 0.1512 | 36.78 | 39.80 | 70.27 | 59.52 | -0.5309 | 10.74 |
| 6-p | 0.1984 | 0.1512 | 37.53 | 40.24 | 71.51 | 60.56 | -0.5136 | 10.50 |

Table 2 shows rating agencies' sample one year loss rates and default rates for investment grade rating letters, to be used as criteria for EL and PD in computations following. Since the standard time horizon for trading book credit risk capital measurements is one year, employing these rates as criteria for counterparty independent haircuts is assuming equivalently that the counterparty will default for sure within the year. The expected loss computed from equation (9) therefore suffices for our purposes. Similarly, to apply S&P's one year default rate targets per rating class, we treat its default rate as the first dollar loss probability in one year horizon and seeks haircuts $h_p$.

Table 2. Moody's idealized expected one-year loss rates (Bielecki 2008) and Standard & Poors' average one-year default rates (S&P 2015).

| Moody's Ratings | Aaa | Aa1 | Aa2 | Aa3 | A1 | A2 | A3 | Baa1 | Baa2 | Baa3 |
|---|---|---|---|---|---|---|---|---|---|---|
| Loss rate(%) | 3.00E-05 | 0.00031 | 0.00075 | 0.00166 | 0.0032 | 0.00598 | 0.02137 | 0.0495 | 0.0935 | 0.231 |
| S&P's Ratings | AAA | AA+ | AA | AA- | A+ | A | A- | BBB+ | BBB | BBB- |
| Default rate(%) | 5.00E-04 | 0.001 | 0.01 | 0.02 | 0.05 | 0.06 | 0.08 | 0.14 | 0.24 | 0.27 |

EL based haircuts results under these three estimations are displayed in Figure 1, where the targeted EL rate is taken from Table 2. Estimation '5-p' and '6-p' produce similar haircuts. '4-p' has systemically lower haircut, due to its poor capture of the sample's skewness and fat tail. For example, to obtain 'Aa2' expected loss, the haircut needs to be 18.5%, 'Aa3' 17%, and 'A' 15.5% close to BASEL's 15% supervisory haircut. For the examples and results following, the full 6-p estimation is applied.

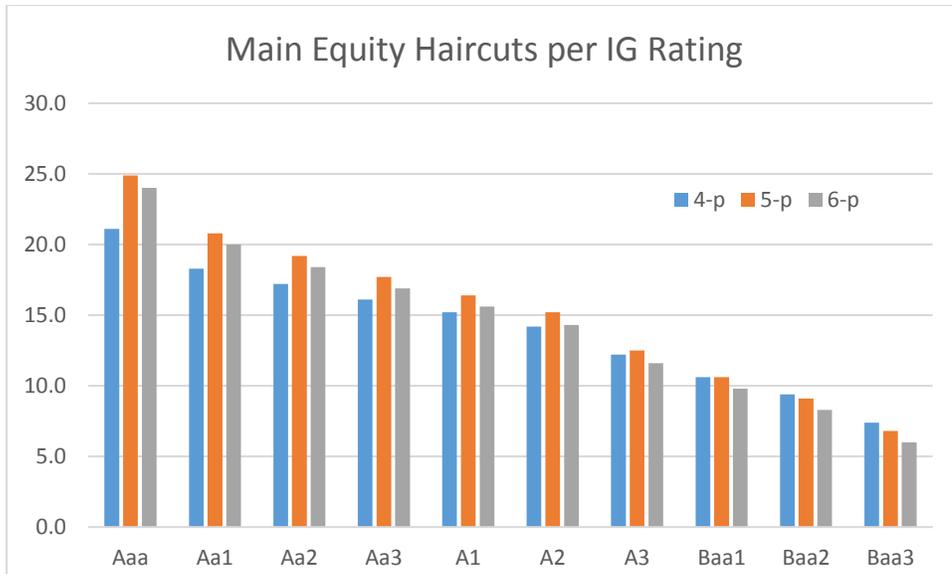

Figure 1. Predicted main equity haircuts vary with rating targets from 'Aaa' to 'Baa3' based on DEJD model estimations of SPX daily prices from 1/2/2008 to 1/2/2013.

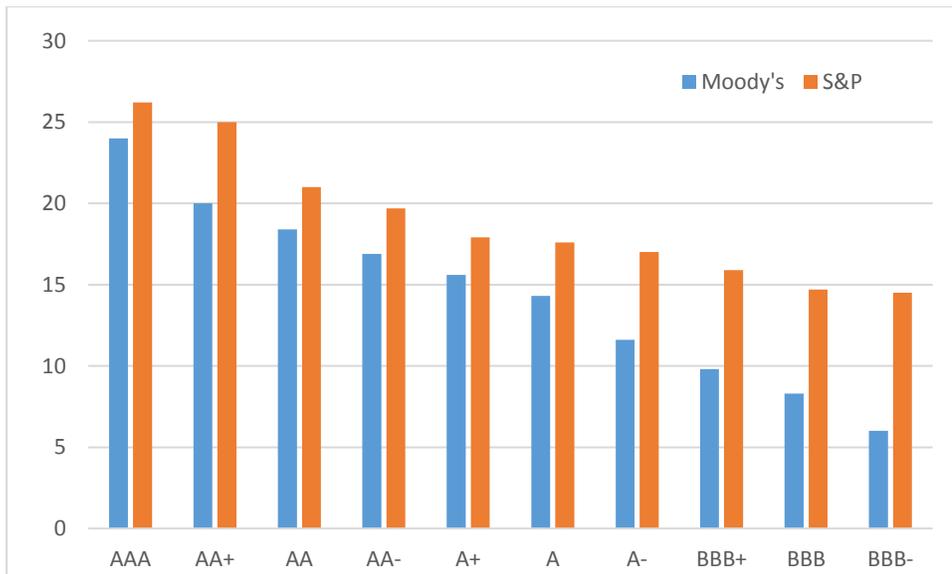

Figure 2. Predicted main equity haircuts (MPR 10 days) targeting hypothetic Moody's and S&P's IG ratings (Table 2).

Figure 2 shows predicted haircuts targeting Moody's and S&P investment grade (IG) credit ratings, i.e., $h_{EL}$ and $h_p$. Note that the default rates in Table 2 are an average of the global corporate default experience between years 1981 to 2014, not necessarily same as S&P's calibrated and official default rates. The hypothetic S&P 'A' and above rating targeted haircuts are about 3~5 points higher than corresponding Moody's rating targeted haircuts. Caution should be taken, however, these default rates and loss rates are examples and not directly comparable against each other. Here our intention is to show that the methodology works for both PD and EL approaches.

**5.2. Securitization debts and liquidity risk**

Non-cash collateral is dominated by fixed income instruments. Unlike main equities, these instruments' market liquidity depth is very limited, except for on-the-run government debts. In practice, a data-driven haircut approach has to rely on non-tradable market indices as proxies to the unobserved data. Because of its fixed maturity, a bond's historical price series (if reliably available) suffers a progressively shortening of the remaining maturity. FSB's haircut framework admits proxies and allows debt securities haircuts to be estimated by putting them into certain residual maturity buckets. For instance, 'AAA' rated US CMBS (commercial mortgage backed securities) bonds are split into 1~5 year and 5~10 year weighted average life buckets.

Repo counterparties wishing to adopt regulatory haircuts often find that the supervisory haircuts table is set too broadly, not distinguishing among many subtypes of assets and risk characteristics of securitized assets. US CMBS, one of the main securitization products seen in repos, is not even found in the table. In fact, there is one broad securitization category. Typical tri-party margin[5] schedules have much finer granularities, with close to a hundred line items offered. With a parametric haircut model in place, more granular haircuts can be computed. Table 3 shows CMBS haircuts predicted using DEJD models of Bank of America/Merrill-Lynch CMBS price indices estimated on daily time series from Jan 2008 to Jan 2013. The maturity upsloping effect is evident as 5~10 year bucket haircuts are higher than those of 1~5 year bucket for all ratings. 'AAA' rated CMBS has much lower haircuts than those of 'AA' and 'A' rated bonds, e.g., 5.08% for 'AAA' 1~5 year bucket as compared to 'AA' at 6.11% and 'A' at 9.14%, when targeting Moody's 'Aa1' rating.

---

[5] A margin ratio is the other side of a haircut, e.g., 20% haircut is equivalent to 125% (=100/(100-20)) margin ratio.

Table 3. Predicted CMBS haircut table per Moody's top three ratings, MPR 10 days. Haircuts from raw data are listed in the last two columns as 99-percentile VaR and ES on 97.5% VaR.

| Rating | Maturity | Aaa | Aa1 | Aa2 | Var99% | ES97.5% |
|---|---|---|---|---|---|---|
| AAA | 1-5y | 6.45 | 5.08 | 4.54 | 4.8 | 5.42 |
| AAA | 5-10y | 15.18 | 12.27 | 11.12 | 11.52 | 12.03 |
| AA | 1-5y | 7.86 | 6.11 | 5.42 | 6.08 | 8.19 |
| AA | 5-10y | 24.94 | 20.35 | 18.51 | 23.49 | 23.91 |
| A | 1-5y | 11.6 | 9.14 | 8.17 | 6.93 | 11.94 |
| A | 5-10y | 24.72 | 20.22 | 18.43 | 26.44 | 24.95 |

The last two columns list raw data haircuts as the standard 10-day 99-percentile VaR or proposed 10-day 97.5-percentile ES per BASEL IV. The estimated models are able to produce haircuts matching closely to those raw haircuts for 1~5y maturity 'AAA' and 'AA'. For 1~5y 'A' VaR and ES have a large deviation, but its average is close to 'Aa1' targeted haircut of 9.14%. Larger differences are seen in 'AA' and 'A' 5~10y maturity, e.g., 'AA' haircut at 20.35% under 'Aa1' target rating while the data has 23.49% haircut from VaR and 23.91 from ES. These differences can be attributed to liquidity risk, yet another FSB's haircut requirement.

Table 4 shows CMBS haircuts targeting 'Aa1' EL (3$^{rd}$ column labeled under 'Aa1' HC) and the additional haircuts (in percentages) when 2% and 5% liquidity premiums (LP) are applied respectively. 2% market liquidity discount can be seen as business usual for non-super senior IG securitized products while 5% is stress. Roughly speaking, these liquidity premiums translate into the same magnitude of additional haircuts. Column 'LP HC' adds '2% LP dHC' to 1~5 year maturity and '5% LP dHC' to 5~10 year maturity and compares with last column's raw haircuts. For 'AA' 5~10y CMBS, the 'Aa1' haircut with 5% liquidity premium is 24.43%, slightly higher than the raw haircut of 23.49%. Note that using prior crisis historical data (Jan 2002 to Jan 2007) only shows 3.44% haircut, while Gorton and Metrick (2012) shows bilateral haircut at 27.5% for 'AA' rated CMBS during the crisis and explains the elevated haircut levels by the liquidity draught facing asset backed securities following the subprime crisis, coupled with sharp increases in experienced volatility and probably anticipated future volatility.

Table 4. Predicted CMBS 10 day haircuts are higher with liquidity risk considered and closer to haircuts observed during the financial crisis.

| Rating | Maturity | Aa1 HC | 2% LP dHC | 5% LP dHC | LP HC | Raw HC (VaR) |
|---|---|---|---|---|---|---|
| AAA | 1-5y | 5.08 | 1.88 | 4.7 | 6.96 | 4.8 |
| AAA | 5-10y | 12.27 | 1.68 | 4.22 | 16.49 | 11.52 |
| AA | 1-5y | 6.11 | 1.82 | 4.56 | 7.93 | 6.08 |
| AA | 5-10y | 20.35 | 1.63 | 4.08 | 24.43 | 23.49 |
| A | 1-5y | 9.14 | 1.84 | 4.6 | 10.98 | 6.93 |
| A | 5-10y | 20.22 | 1.65 | 4.12 | 24.34 | 26.44 |

**5.3. Corporate debts and idiosyncratic adjustments**

The last major asset class demonstrated is US IG corporate bonds. Table 5 shows single 'A' rated bonds with residual maturities of 1~5, 5~10, and 10~15 years. Comparing to BASEL III supervisory haircuts for 20% risk weighted wholesale issuers at 4% for residual maturity of 1~5 years and 8% for 5+ years, 'Aaa' is the rating letter that gets closest to the supervisory haircuts. 'Aa1' is not far behind. Differences between 'Aa1' targeted haircuts and the raw haircuts are in the same scale as bond bid/ask spreads, which can be added through the liquidity discount as shown previously.

Table 5. Predicted 'A' rated US corporate bond haircuts per top three ratings with comparison to raw haircuts, MPR=10 days, using Bank of America/Merrill-Lynch US corporate bond price indices from January 2008 to January 2013.

| Rating | Maturity | Aaa | Aa1 | Aa2 | Raw HC |
|---|---|---|---|---|---|
| A | 1-5y | 3.87 | 2.98 | 2.64 | 4.25 |
| A | 5-10y | 6.49 | 5.19 | 4.68 | 6.43 |
| A | 10-15y | 6.15 | 5.02 | 4.59 | 5.6 |

Once a proxy historical price series (such as the ones in Table 5) is chosen and justified, the collateral haircut is trivially given via the VaR approach. A parametric model could be separately formed by estimating the DEJD model from the proxy and predicts haircuts. But, if its sole use is to reproduce collateral haircuts, it obviously does not add any value. The parametric model is useful, however, in that it facilitates a meaningful sensitivity analysis of the collateral haircut, as shown in Table 6. The sens to up jump rate and down jump rate $(\eta_u, \eta_d)$ are asymmetric as expected, for haircut measures one sided loss and depends on dump jumps rather than up jumps. With 1% shift in volatility, haircut increases by an amount of 0.33%., or roughly 3% shift in volatility will lead to a haircut increase of 1%. Given increased market volatility, haircut deltas or adjustment can then be computed and added to applicable haircuts. Scenario analysis and stress test can be conducted as desired.

Moreover, the parametric model allows idiosyncratic factors to be incorporated as adjustments. When model estimation is performed based on an index return, index constituents are averaged out. For a long running index, constituents credit quality and market performances will diverge as it ages. The resulted haircuts from a data-driven index-proxy approach could be too generous for some subgroups of the index portfolio. So far, practitioners don't have a tool to adjust this aging effect. With a parametric model at hand, one could create a delta model for the deteriorating credits. For instance, the down jump size distribution parameter $\eta_d$ can be adjusted down to reflect larger jump sizes, or the down jump intensity (through $\lambda$ and $1-p$) can be increased, or the volatility $\sigma_a$ can be hiked. On a first order basis, the sens table thus could be used to incorporate idiosyncratic risk characteristics within a proxy index or portfolio.

Table 6. Haircuts' sensitivities to DEJD model parameters shown for 'A' rated 5 to 10 year maturity corporate bond. MPR 10 days, $g=0$. Shifts are based on $(\mu, \sigma_a, \lambda_u, \lambda_d, \eta_u, \eta_d) = (0.0729, 0.0525, 13.82, 31.90, 212.6, 225.6)$.

| Shifts | Aaa | Aa1 | Aa2 |
|---|---|---|---|
| $\mu$+1% | -0.03 | -0.04 | -0.04 |
| $\sigma_a$ +1% | 0.37 | 0.34 | 0.32 |
| $\lambda_u$-1 | 0.01 | 0.01 | 0 |
| $\lambda_d$+1 | 0.07 | 0.04 | 0.04 |
| $\eta_u$ + 10 | 0.01 | 0 | 0 |
| $\eta_d$ − 10 | 0.26 | 0.2 | 0.18 |

These predicted haircuts of major classes of collateral assets, although not comprehensive, do seem to indicate that the expected loss matching a top credit quality scale (e.g., 'Aa1/Aa2') satisfies FSB's quantitative standards and produces numbers generally in good agreement with supervisory haircuts. Since S&P's 'AAA' or Moody's 'Aaa' rated corporates are numbered and there is basically no default history to study their default frequency or expected loss, for haircut purposes, we recommend the next top two ratings, 'AA+' ('Aa1') or 'AA' (Aa2)[6].

### 5.4. Parametrization stability and model risk

To get some sense of estimation stability, we conduct quarterly rolling SPX model estimations starting from 2005 Q1 to 2009 Q4. For each quarter, a 5 year historical daily return series is taken, for example, the 2005 Q1 sample covers data series from 1/1/2005 to 1/1/2010. Note that the height of the financial crisis could be taken as the stress period from 7/1/2008 to 7/1/2009, so the last two estimations (on 2009 Q3 and 2009 Q4) are already out of the stress period. Figure 3 shows model computed standard deviations and haircuts when EL targets at 'Aa2' ratings. Drops for 2008 Q3/Q4 samples, because the remaining of the 5 year time series lies mostly out of the stress period, although they start deep in the crisis. Estimated models' skewness are generally in line with data sample's skewness.

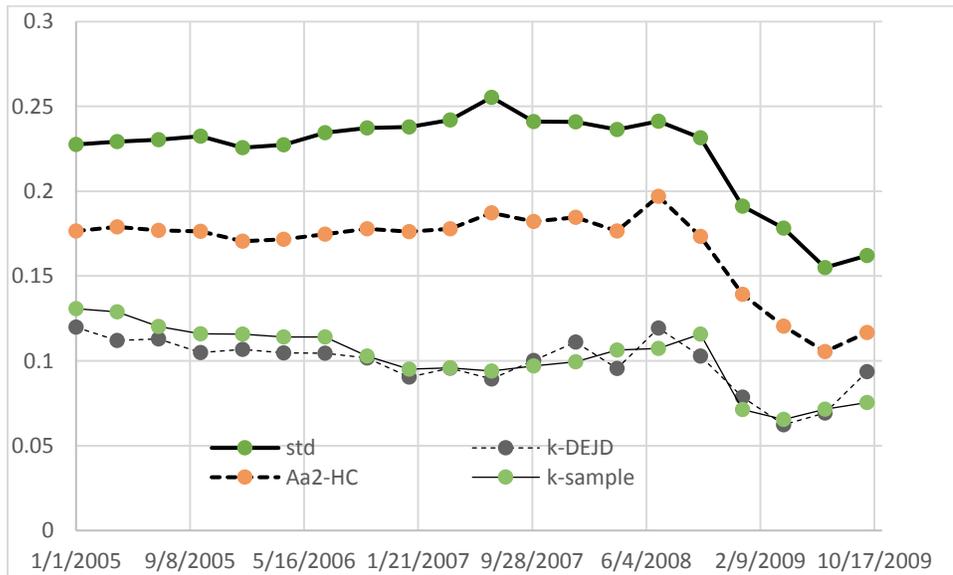

Figure 3. Estimated sample standard deviation and 'Aa2' haircuts change as the 5 year SPX series rolls from 1/1/2005 to 10/1/2009. Also shown are sample kurtosis (labeled as 'k-sample') and model kurtosis ('k-model'), scaled by 0.01.

---

[6] Banks' wholesale credit risk measurement methodologies normally combine 'AAA' ('Aaa') and 'AA+' (Aa1) into one top credit quality rating.

Haircuts exhibited in Figure 3 are stable and in sync with the estimated volatility. The jump parameters are shown in Figure 4. The volatile time comes in 2008, in sync with the financial crisis in a higher drive. Approaching the end of sample series, intensities for both up and down jumps are weaker and average jump sizes also decrease (the average up jump size is reciprocal of $\eta_u$), as the index has a much less volatile period following Q3 2009. The overall behavior is expected and relative stable.

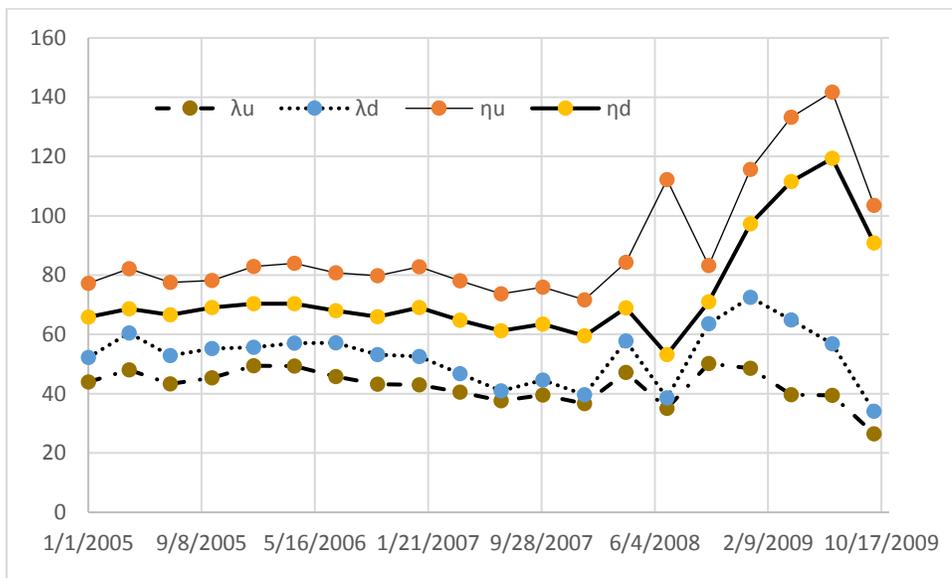

Figure 4. Estimated S&P 500 DEJD jump parameters during 1/1/2005 to 10/1/2009.

As discussed earlier, overly depending on historical data is a source of model risk, e.g., the subprime mortgage bonds. An alternative to estimated models is a model calibrated to options markets. Table 7 shows haircuts under each of two calibrated MEM models for SPX index: Cai-Kou (2011) is calibrated to SPX European options, while CKL 2014 (Cai et al 2014) has more weights towards SPX option smiles. The last column shows haircuts needed to produce no-loss with confidence level of 99.9. Cai-Kou2011 model is in line with the 15% supervisory haircut at 'Aa2' rating, 'Aa3', and 99.9-percentile. The option implied CKL-2014 model overshoots haircuts. This reflects the danger of exclusively calibrating to deep out-of-the-money put options, as these options are known of consisting of significant liquidity premiums, not reflecting underlying securities or indices' liquidity.

Goodness of fit tests for DEJD model has been done for SPX in Ramezani and Zeng (2007). Judging from the skewness and kurtosis as seen from the rolling SPX samples above, we can conclude that SPX is fit satisfactory, without having to repeat those tests. A separate assessment of goodness of fit for corporate bonds and securitization debts will be a worthwhile effort in the future. As appropriate, once model risk is determined and quantified to be substantial, a haircut add-on could be levied to compensate it.

Table 7. SPX haircuts (MPR=10 days) under two implied risk-neutral DEJD models, compared with raw haircut of 14.44% and regulatory haircut of 15%.

|  | Aa1 | Aa2 | Aa3 | 99.9% |
| --- | --- | --- | --- | --- |
| CKL-2014 | 34.48% | 31.48% | 28.66% | 25.6% |
| Cai-Kou2011 | 17.86% | 15.93% | 14.15% | 13.4% |

## 6. Conclusion

As non-cash collateral under CSAs is used to mitigate counterparty credit risk, a separation of collateral risk from OTC derivatives exposure risk is desired. This is the rationale behind BASEL III and FSB's enhanced haircut framework that explicitly requires counterparty independent haircuts and establishes qualitative and quantitative standards on the prevailing data-driven haircut approach. A data-driven approach such as the VaR model, however, is subject to limitations in terms of data availability, reliability, and flexibility.

We propose an alternative, complementary parametric haircut model aiming at conducting haircut sensitivity analysis, capturing liquidity risk and idiosyncratic risk, and incorporating useful information from related markets. For example, proxy data series are often used in practice, but a proxy fails to capture a specific bond's recent price behavior and credit deterioration when it becomes evidently idiosyncratic. The parametric model allows us to stress volatility or downward jump magnitude or probability to attempt to reflect deterioration of credits as compared to its proxy.

To that end, haircuts definitions are extended so that the credit risk profile of a haircut non-cash collateral asset achieves certain performance criteria, such as minimum expected loss or probability of default given certain high rating targets (e.g. S&P's 'AA+' or Moody's 'Aa2') or minimal economic capital (e.g. one year, 99.9-percentile credit risk VaR). Double-exponential jump-diffusion process is used to model the asset return and produce the market risk exposure. The haircut model can be extended to study counterparty dependent haircuts (e.g. repo haircuts), with wrong way risk consideration for instance.

Preliminary results show that estimated DEJD models with a stress period are able to produce haircut levels consistent with collateral haircuts for equities, corporate bonds and securitized products as typically seen in CSAs and BASEL. A built measure of market liquidity together with expected increases of future volatility can explain the haircut increases during the financial crisis. The model can also help design CSA haircuts and expand the supervisory haircut table with much finer granularity in terms of its product coverage.

Haircuts are intimately related to the loss side of the tail distribution. A study of haircut is necessarily a study on tail behavior. The DEJD model is known to provide a reasonable fit for stock indices. Its goodness of fit tests for corporate bonds and securitization debts will be left for future research. Application of MEM as an extension of DEJD model will be explored with the view that its enhanced skew and tail capturing ability might be needed for securitized debts structural linkage exists. Also interesting is to see stochastic volatility jump diffusion models' performance for haircut purposes.

## Appendix: Laplace Inversion Procedures

Cai et al (2014) proposes two sided Laplacian inversion algorithm. To solve for $f_{X(t)}$,

$$f(t) = f_A(t, \sigma, C, N) + e_T(t, \sigma, C, N) - e_D(t, \sigma, C),$$

$$f_A(t, \sigma, C, N) = \frac{exp(\sigma t) L_f(\sigma)}{2(|t|+C)}$$

$$+ \frac{exp(\sigma t)}{|t|+C} \sum_{k=1}^{N} [(-1)^k Re(exp(-\frac{k\pi i C sgn(t)}{t+C sgn(t)}) L_f(\sigma + \frac{k\pi i}{t+C sgn(t)}))] \qquad (A1)$$

where $f_A$ provides an accurate approximation of $f(t)$ when the truncation error $e_T$ and the discretization error $e_D$ are small. $\sigma$ is a number in the ROAC, $C>0$ a shift constant to control $e_D$, and $N>0$ the number of truncation terms to control $e_T$.

The truncation error $e_T$ is bounded as,

$$|e_T(t, \sigma, C, N)| \leq \frac{\varsigma(\sigma) e^{\sigma t}}{\pi \xi \rho^{(1-\beta)/\xi}} \Gamma(\frac{(1-\beta)}{\xi}, \rho(\frac{\pi}{|t|+C} N)^{\xi}), \qquad (A2)$$

where $\Gamma(a,b)$ is the upper incomplete gamma function of order $a$ and lower bound $b$. For the inversion of pdf Laplacian $L_f$, these parameters are listed below,

$$\beta = 0, \rho = \frac{1}{2}\sigma_a^2 t, \xi = 2, \zeta(\sigma) = e^{tG(-\sigma)}, \qquad (A3)$$

For *cdf* $L_F$, use $\beta=1$. For European put options $L_P$, $\beta=2$ and $\zeta(-\sigma-1)$ can be easily shown, following Cai et al (2014)'s derivation.

We propose a different set of parameters so that the relative errors can be determined for all three inversions, equations (A4.a, A4.b, and A4.c) for pdf, cdf and put respectively.

$$\beta = 0, \rho = \frac{1}{2}\sigma_a^2 t, \xi = 2, \zeta(\sigma) = e^{tG(-\sigma)}, \qquad (A4.a)$$

$$\beta = 0, \rho = \frac{1}{2}\sigma_a^2 t, \xi = 2, \zeta(\sigma) = \frac{e^{tG(-\sigma)}}{|\sigma|}, \qquad (A4.b)$$

$$\beta = 0, \rho = \frac{1}{2}\sigma_a^2 t, \xi = 2, \zeta(\sigma) = \frac{e^{tG(\sigma+1)}}{(\sigma+1)^2}, \qquad (A4.c)$$

Noting that $\varsigma(\sigma)e^{\sigma t}$ term in equation (A2) also appears in $f_A$'s leading term, while the rest of variables in (A2) does not depend on which Laplacian is being converted, the relative errors of $e_T$ to $f_A$ is therefore independent, so that the same $N$ can be used for all three inversions.

The discretization error derived in Cai et al (2014) is more complex. One can choose a sufficiently large C to allow fast decays. In Cai et al (2014), different *C* and *N* being used when calculating *pdf*, *cdf*, and call option. This is disadvantageous, the expected loss and tail cdf of loss (which relate to put option and cdf of $X_t$ respectively) need to be calculated at the same time, especially when the return dynamics is coupled with counterparty credit dynamics (Lou 2016b). The moneyness normalized strike is then critical to compute them efficiently under the same set of parameters to employ adaptive error control that ensures sufficient accuracy.